\documentstyle[fullpage]{article}

\newcommand{\la}{\lambda}
\newcommand{\si}{\sigma}

\title{The $r$-matrix structure of the Euler-Calogero-Moser model}
\author{E. Billey $^*$ \and J. Avan $^*$ \and O. Babelon \thanks{L.P.T.H.E.
Universit\'e Paris VI (CNRS UA 280), Box 126, Tour 16,
$1^{er}$ \'etage, 4 place Jussieu, F-75252 PARIS CEDEX 05}}
\date{December 1993}

\begin{document}

\begin{titlepage}
\renewcommand{\thepage}{}
\maketitle
\vspace{2cm}
\begin{abstract}
We construct the $r$-matrix for the generalization of the Calogero-Moser
system introduced by Gibbons and Hermsen. By reduction procedures we
obtain the $r$-matrix for the $O(N)$ Euler-Calogero-Moser model and for the
standard $A_N$ Calogero-Moser model.
\end{abstract}

\vfill

PAR LPTHE 93-55

\end{titlepage}
\renewcommand{\thepage}{\arabic{page}}

\section{Introduction}

The Calogero-Moser model seems to occupy a particular place in the world of
integrable systems. On one hand, its complete solution is known
\cite{C,D,Po,BHV} even at the quantum level, but on the other hand, the
techniques which have now become standard --- in particular the $r$-matrix
formalism --- did not seem to apply to it.

The beginning of an answer to this problem was given in \cite{AT1} where the
classical $r$-matrix was calculated and found to be a dynamical object,
which is a non standard feature. This result was later extended to the
elliptic case in \cite{S1,BS}.

While the general theory of constant classical $r$-matrices is now well
developped, this is not so for the dynamical ones and we still are in the
phase of producing examples. Dynamical $r$-matrices appear when we consider
higher Poisson structures \cite{A,S2}. They also appear in non linear
$\si$-models \cite{BFLS,Ma}, or in the theory of Sine Gordon solitons
\cite{BB}. Finally one should  mention an interesting
example of dynamical $r$-matrix for the hyperbolic Gaudin model \cite{EEKT}.

In this paper we will consider the generalization of the Calogero-Moser model
introduced by Gibbons and Hermsen \cite{GH}. This class of models has recently
received some attention in relation with the evolution of energy levels with
respect to the parameter $t$ for a perturbated system of Hamiltonian
$ H = H_0 + t V $ \cite{Pe,Y,NL}.

Let us introduce a set of dynamical variables $ (q_i,p_i)_{i=1 \cdots N} $ and
$ (f_{ij})_{i,j=1 \cdots N} $ together with the Poisson brackets
\begin{eqnarray}
  \{ p_i, q_j \} & = & \delta_{ij} \\
  \label{var}
  \{ f_{ij}, f_{kl} \} & = &  \delta_{jk} \ f_{il}
                                 - \delta_{li} \ f_{kj}
\end{eqnarray}
\noindent and the Hamiltonian
\begin{equation}
  \label{H}
  H = \frac{1}{2} \sum_{i=1}^N p_i^2
    - \frac{1}{2} \sum_{_{\ i \neq j}^{i,j=1}}^N
      \frac{f_{ij}f_{ji}}{(q_i-q_j)^2} \ .
\end{equation}

In order to have a non degenerate Poisson bracket it is assumed that the
$ (f_{ij})_{i,j=1 \cdots N} $ are restricted to a symplectic submanifold
of (\ref{var}).

Notice furthermore that the quantities $ (f_{ii})_{i=1 \cdots N} $ Poisson
commute with the Hamiltonian; we can therefore reduce the flows to the
surfaces $ (f_{ii} = \mbox{constant})_{i=1 \cdots N} $. These models  were
shown to be integrable precisely on these surfaces and our purpose is to
compute their $r$-matrices.

The plan of the paper is as follows : after a short review on the $r$-matrix
formalism in section 2, we find in section 3 the Poisson structure of the
Lax operator which implies the integrability property of the model on the
manifolds $ (f_{ii} = \mbox{constant})_{i=1 \cdots N} $. In section 4 we
use the previous result to obtain the $r$-matrix structure for the $O(N)$
Euler-Calogero-Moser model introduced by Wojciechowski \cite{W}. This is
achieved by using the mean procedure introduced in \cite{Mi,FR,AT2}. Finally in
section 5 we again use the results of section 3 to present a new construction
of the $r$-matrix of the standard Calogero-Moser model computed for the first
time in \cite{AT1} and obtained here by Hamiltonian reduction from the model
(\ref{H}).

\section{Dynamical $r$-matrices}

In the Lax representation of a dynamical system, the equations of motion are
written in the form
\begin{equation}
  \label{Lax}
  \dot{L} = [L,M]
\end{equation}
\noindent where $L$ and $M$ take values in a Lie algebra $\cal{G}$ with
basis $ \{ X_{\mu} \}. $

Eq.(\ref{Lax}) implies that the quantities $tr(L^n)$ are conserved. Liouville
integrability requires their involution under the Poisson brackets.
It was shown in \cite{BV} that this property $ \{tr(L^m),tr(L^n)\}
= 0 $ is equivalent to the existence of an $r$-matrix.

If we set
$$ L_1 = L \otimes 1 $$
$$ L_2 = 1 \otimes L $$
\noindent this condition reads
\begin{equation}
  \label{pb}
  \{L_1,L_2\} = [r_{12},L_1] - [r_{21},L_2]
\end{equation}
\noindent where
$$ r_{12} = \sum r_{\mu \nu} X_{\mu} \otimes X_{\nu} \ \ \ \mbox{and} \ \ \
   r_{21} = \sum r_{\mu \nu} X_{\nu} \otimes X_{\mu}. $$

In general the $r$-matrix is a dynamical object and possesses no special
symmetry. Let us recall here some of its properties.

\begin{enumerate}
\item The Lax operator $L$ together with the $r$-matrix completely describe the
dynamical system: if we choose $\frac{1}{n}tr(L^n)$ as a Hamiltonian, the
equations of motion take the Lax form with $M=tr_2(r_{12} L_2^{n-1}).$
\item The left-hand side of eq.(\ref{pb}) being a Poisson bracket, the
Jacobi identity implies a constraint on $r$. It is easy to check that it
takes the form
\begin{equation}
  \label{Jacob}
  \begin{array}{c}
   [L_1,[r_{12},r_{13}]+[r_{12},r_{23}]+[r_{32},r_{13}]+ \{L_2,r_{13}\}-
       \{L_3,r_{12}\}]  \\
   + \ \mbox{cyclic permutations} = 0 \ .
  \end{array}
\end{equation}
It is interesting to note that the display of indices in (\ref{Jacob}) is not
the one which appears in the usual Yang-Baxter equation. We recover it when
the $r$-matrix is antisymmetric.
\item Since the existence of an $r$-matrix is equivalent to the involution
property of the eigenvalues of $L$, the form of eq.(\ref{pb}) is not modified
by a conjugation of the Lax matrix. Indeed if
$ L' = g^{-1} L g $ then
$$ \{ L'_1,L'_2\ \} = [r'_{12},L'_1] - [r'_{21},L'_2] $$
with
\begin{equation}
  \label{newpb}
  r'_{12} = g_1^{-1} g_2^{-1} \left ( r_{12}
  - \{ g_1,L_2 \} g_1^{-1} + \frac{1}{2} [u_{12},L_2] \right ) g_1 g_2
\end{equation}
\noindent where $ u_{12} = \{ g_1,g_2 \} g_1^{-1} g_2^{-1} $~.
\item Finally we recall that the $r$-matrix of a given Lax matrix is not
unique. In particular if we introduce a symmetric matrix $\tau_{12} =
\tau_{21}$ of ${\cal G} \otimes {\cal G}$, the change
\begin{equation}
  \label{ambig}
   r_{12} \longrightarrow r_{12} + [\tau_{12},L_2]
\end{equation}
\noindent does not affect the equation (\ref{pb}).
\end{enumerate}

\section{The Poisson structure for the $ sl(N) $ model}

The Lax matrix of the system (\ref{H}) is
\begin{equation}
  \label{sln}
  L(\la) = \sum_{i=1}^N \left ( p_i - \frac{f_{ii}}{\la} \right ) \ e_{ii}
                               - \sum_{_{\ i \neq j}^{i,j=1}}^N
                               \left ( \frac{1}{q_i-q_j} + \frac{1}{\la}
                               \right ) f_{ij} \ e_{ij} \ .
\end{equation}
where $ (e_{ij})_{kl} = \delta_{ik} \delta_{jl}.$

The Poisson brackets of the elements of the Lax matrix (\ref{sln}) can be
recast in the form
\begin{equation}
  \label{ps}
  \{ L_1(\la) , L_2(\mu) \} = [r_{12}(\la,\mu),L_1(\la)] -
                              [r_{21}(\mu,\la),L_2(\mu)] -
                              \sum_{_{\ i \neq j}^{i,j=1}}^N
                              \frac{f_{ii} - f_{jj}}{( q_i - q_j )^2}
                              \ e_{ij} \otimes e_{ji}
\end{equation}
where
\begin{equation}
   \label{r12}
   r_{12}(\la,\mu) = - \frac{\cal C}{\la - \mu}
                     - \sum_{_{\ i \neq j}^{i,j=1}}^N \frac{1}{q_i - q_j} \
                       e_{ij} \otimes e_{ji}
\end{equation}
\noindent and ${\cal C}$ is the Casimir element of $sl(N)$
$$ {\cal C}  =  \sum_{i,j=1}^N e_{ij} \otimes e_{ji}.$$
Some comments are in order:
\begin{enumerate}
\item Eq.(\ref{ps}) holds for the non reduced dynamical system
(\ref{H}) and it is known that $ trL^n(\la) $ are not in involution for this
system. This fact is responsible for the additional term in eq.(\ref{ps})
compared to eq.(\ref{pb}). Indeed we have
$$ \{trL^n(\la) , trL^m(\mu) \} = n \ m \sum_{_{\ i \neq j}^{i,j=1}}^N
   \frac{f_{ii}-f_{jj}}{(q_i-q_j)^2} \ [L^{n-1}(\la)]_{ij}
   \ [L^{m-1}(\mu)]_{ji}.$$
\item The constraints $f_{ii}$ generate on $L(\la)$ a conjugation by a diagonal
matrix. Hence $trL^n(\la)$ Poisson commute with $f_{ii}$. It follows that one
can compute their reduced Poisson bracket on the manifold $ (f_{ii} =
\mbox{constant})_{i=1 \cdots N} $. This amounts to setting $ f_{ii} =
\mbox{constant} $ in eq.(\ref{ps}) and therefore $trL^n(\la)$ are in
involution.
\item Since the quantities $trL^n(\la)$ commute on the reduced phase space,
$L(\la)$ has an $r$-matrix. However its explicit computation requires some
care since $L(\la)$ is not a function on the reduced phase space.
\end{enumerate}
In the following two sections we give two examples of reductions where we
can explicitely obtain the $r$-matrix, starting from the initial
structure (\ref{ps}).

\section{The $r$-matrix for the $ O(N) $ model}

We now consider the following model, introduced by Wojciechowski
in \cite{W}. The dynamical variables are $ (q_i,p_i)_{i=1 \cdots N} $ and
antisymmetric $ (h_{ij} = - h_{ji})_{i,j=1 \cdots N} $. The Poisson bracket
is now:
\begin{eqnarray}
   \{ p_i, q_j \} & = & \delta_{ij} \\
  \label{newvar}
  \{ h_{ij}, h_{kl} \} & = &  \frac{1}{2} \ (
                              \delta_{il} \ h_{jk} +
                              \delta_{ki} \ h_{lj} +
                              \delta_{jk} \ h_{il} +
                              \delta_{lj} \ h_{ki}  ) \ .
\end{eqnarray}

The Hamiltonian of the system reads
$$ H = \frac{1}{2} \sum_{i=1}^N p_i^2  + \frac{1}{2}
                   \sum_{_{\ i \neq j}^{i,j=1}}^N
                    \frac{h_{ij}^2}{(q_i-q_j)^2} \ . $$

As shown in \cite{W} the equations of motion can be written in the form
$$ \dot{L}(\la) = [L(\la),M(\la)] $$
where the Lax pair $(L,M)$ is given by
\begin{eqnarray}
  \label{L}
  L(\la) & = & \sum_{i=1}^N p_i \ e_{ii}
                              - \sum_{_{\ i \neq j}^{i,j=1}}^N
                                \left ( \frac{1}{q_i-q_j} + \frac{1}{\la}
                                 \right ) h_{ij} \ e_{ij} \\
  M(\la) & = & \sum_{_{\ i \neq j}^{i,j=1}}^N
               \frac{h_{ij}}{(q_i-q_j)^2} \ e_{ij} \ ,
\end{eqnarray}
and the Hamiltonian is $ H = \frac{1}{2} \int \frac{d\la}{2i\pi\la}
                                                trL(\la)^2 $.

In \cite{W} this system was proved to have Poisson commuting Hamiltonians
$trL^n(\la)$. It follows that this model admits an $r$-matrix, which we now
calculate.

First of all we remark that the Lax matrix of the $O(N)$ model is obtained
from the Lax matrix of the $sl(N)$ model as
\begin{equation}
  L^{O(N)}(\la) = \frac{1}{2} \ (1-\si) L^{sl(N)}(\la)
\end{equation}
where $\si$ is the following involutive automorphism of the loop algebra:
\begin{equation}
  \si : \la^n e_{ij} \longrightarrow - (-\la)^n e_{ji}.
\end{equation}
In this average procedure
\begin{equation}
  h_{ij} = \frac{1}{2} (f_{ij} - f_{ji})
\end{equation}
and the Poisson structure (\ref{var}) of the $sl(N)$ model becomes the
structure (\ref{newvar}).

It follows immediately that
\begin{eqnarray}
 \{ L_1^{O(N)}(\la) , L_2^{O(N)}(\mu) \} = \frac{1}{4} \
       (1- \si \otimes 1 - 1 \otimes \si
      \! \! \! \! & + & \! \! \! \! \si \otimes \si ) \cdot \nonumber \\
     {\Big(} [r_{12}(\la,\mu),L_1^{sl(N)}(\la)] \! \! \! \!
                      & - & \! \! \! \! [r_{21}(\mu,\la),L_2^{sl(N)}(\mu)]
                       - \sum_{_{\ i \neq j}^{i,j=1}}^N
                         \frac{f_{ii} - f_{jj}}{( q_i - q_j )^2}
                         \ e_{ij} \otimes e_{ji} {\Big)}.
\end{eqnarray}
We now remark that
\begin{equation}
  ( 1- \si \otimes 1 - 1 \otimes \si + \si \otimes \si )
    \sum_{_{\ i \neq j}^{i,j=1}}^N
    \frac{f_{ii} - f_{jj}}{( q_i - q_j )^2}
    \ e_{ij} \otimes e_{ji} = 0.
\end{equation}
 Moreover since
\begin{equation}
\si \otimes \si \ r_{12}(\la,\mu) = - r_{12}(\la,\mu)
\end{equation}
we finally get an explicit $r$-matrix structure for $L^{O(N)}(\la)$:
\begin{equation}
  \{ L_1^{O(N)}(\la) , L_2^{O(N)}(\mu) \} =
                   [r_{12}^{O(N)}(\la,\mu),L_1^{O(N)}(\la)]
                 - [r_{21}^{O(N)}(\mu,\la),L_2^{O(N)}(\mu)]
\end{equation}
with
\begin{equation}
  r_{12}^{O(N)}(\la,\mu) = \frac{1}{2}(1+\si \otimes 1) r_{12}^{sl(N)}(\la,\mu)
\end{equation}
or explicitely
\begin{eqnarray}
  \label{r}
  r_{12}^{O(N)}(\la,\mu) = - \frac{\la}{\la^2-\mu^2}
                              \sum_{i=1}^N e_{ii} \otimes e_{ii}
                          & - & \frac{1}{2} \sum_{_{\ i \neq j}^{i,j=1}}^N
                                \left ( \frac{1}{q_i-q_j} + \frac{1}{\la+\mu}
                                \right ) e_{ij} \otimes e_{ij} \nonumber \\
                          & - & \frac{1}{2} \sum_{_{\ i \neq j}^{i,j=1}}^N
                                \left ( \frac{1}{q_i-q_j} + \frac{1}{\la-\mu}
                                \right ) e_{ij} \otimes e_{ji}.
\end{eqnarray}
This is an example of application of the well known mean procedure
\cite{Mi,FR,AT2}.

It is interesting to see how the Jacobi identity (\ref{Jacob}) is fulfilled for
this dynamical $r$-matrix (\ref{r}). We find that
$$ [r_{12},r_{13}]+[r_{12},r_{23}]+[r_{32},r_{13}] =
   \frac{1}{2} \sum_{_{\ i \neq j}^{i,j=1}}^N
         \frac{1}{(q_i - q_j)^2} e_{ij} \otimes
         \left [ ( e_{ij} + e_{ji} ) \otimes ( e_{ii} - e_{jj} )
         - ( e_{ii} - e_{jj} ) \otimes ( e_{ij} + e_{ji} ) \right ] \ .$$

It is then easy to check that
\begin{equation}
  \label{gYB}
  [r_{12},r_{13}]+[r_{12},r_{23}]+[r_{32},r_{13}]+ \{L_2,r_{13}\}-
       \{L_3,r_{12}\} = 0.
\end{equation}
The remarkable feature of this equation is the vanishing of its right-hand
side. This is in contrast with other cases \cite{S1,EEKT} where eq.(\ref{gYB})
has a non vanishing right-hand side, albeit of a special form.

\section{A new construction of the $r$-matrix for the Calogero model}

We now show how the dynamical $r$-matrix for the usual Calogero model
\cite{AT1,S1,BS} stems from eq.(\ref{ps}) through a procedure of Hamiltonian
reduction.

Let us restrict oneselves to the symplectic manifold
\begin{equation}
  f_{ij} = \xi_i \ \eta_j
\end{equation}
with $ \{\xi_i,\xi_j\}=0 , \ \{\eta_i,\eta_j\}=0 ,
                       \ \{\xi_i,\eta_j\}= - \delta_{ij}. $
The Poisson brackets of $f_{ij}$ are indeed given by (\ref{var}).

On this manifold we have a symplectic action of an Abelian Lie group
\begin{equation}
  \xi_i \longrightarrow \la_i \ \xi_i , \ \ \
 \eta_i \longrightarrow \la_i^{-1} \eta_i.
\end{equation}
The Hamiltonian (\ref{H}) is invariant under this action and we can apply the
method of Hamiltonian reduction. The infinitesimal generator of this action is
$$ H_\epsilon = \sum_{i=1}^N \epsilon_i f_{ii}. $$
The reduction is performed by first fixing the momentum. We choose:
$$ f_{ii}=\alpha. $$
The isotropy group $G_{\alpha}$ of $\alpha$ is the whole group itself since
it is Abelian and we still have to quotient out the action of this group. At
the end of the procedure, the $2N$ degrees of freedom $(\xi_i,\eta_i)_{i=1
\cdots N}$ are eliminated, leaving as reduced Hamiltonian (\ref{H}) the
Hamiltonian of the usual Calogero-Moser model
$$ H_{\rm Cal} =  \frac{1}{2} \sum_{i=1}^N p_i^2
    - \frac{1}{2} \ \alpha^2 \sum_{_{\ i \neq j}^{i,j=1}}^N
      \frac{1}{(q_i-q_j)^2} \ . $$
In order to perform the reduction at the level of the Lax matrix we remark
that the group acts on $L$ as conjugation by the matrix $ {\rm diag}(\la_i)_
{i=1 \cdots N}.$ We now observe that the matrix
$$ L_{\rm Cal} = {\rm diag}(\xi_i^{-1}) \ L \ {\rm diag}(\xi_i) $$
is invariant under $G_{\alpha}$ and we can therefore compute the Poisson
brackets of its matrix elements safely.
\begin{equation}
  L_{\rm Cal} = \sum_{i=1}^N \left ( p_i - \frac{\alpha}{\la} \right ) e_{ii}
              - \alpha \sum_{_{\ i \neq j}^{i,j=1}}^N
                    \left ( \frac{1}{q_i-q_j} + \frac{1}{\la} \right ) e_{ij}.
\end{equation}
The $r$-matrix of this $L$ matrix is now obtained from eq.(\ref{ps}) and
(\ref{r12}) by applying the conjugation formula~(\ref{newpb}).

When setting $f_{ii}=\alpha$ the extra term in eq.(\ref{ps}) vanishes, leaving
us with the $r$-matrix
$$ r_{12}^{\rm Cal}(\la,\mu) = r_{12}(\la,\mu)
                    - \frac{1}{\mu} \sum_{i=1}^N e_{ii} \otimes e_{ii}
                    + \sum_{_{\ i \neq j}^{i,j=1}}^N \left (
                      \frac{1}{q_i-q_j} - \frac{1}{\mu} \right )
                      e_{ii} \otimes e_{ji}$$
which gives
\begin{equation}
  r_{12}^{\rm Cal}(\la,\mu) = - \frac{\la}{\mu(\la-\mu)}
                                \sum_{i=1}^N e_{ii} \otimes e_{ii}
                              - \sum_{_{\ i \neq j}^{i,j=1}}^N \left (
                      \frac{1}{q_i-q_j} + \frac{1}{\la-\mu} \right )
                      e_{ij} \otimes e_{ji}
                              + \sum_{_{\ i \neq j}^{i,j=1}}^N \left (
                      \frac{1}{q_i-q_j} - \frac{1}{\mu} \right )
                      e_{ii} \otimes e_{ji}.
\end{equation}
This $r$-matrix does not coincide with the $r$-matrix obtained in
\cite{AT1,S1,BS}, however it is easily related to it by a transformation
(\ref{ambig}) with $\tau_{12}=\frac{1}{2} \sum_{i=1}^N e_{ii} \otimes e_{ii}$.

\section{Conclusion}

The class of models introduced by Gibbons and Hermsen \cite{GH} turns out to
be a particularly interesting generalization of the Calogero model. In
particular their $r$-matrix structure is rather elegant and is general enough
to allow the construction of a host of new models by various reduction
procedures. Generalization to trigonometric and elliptic potentials is
possible and will be developped further.

\bigskip

\bigskip

\noindent{\Large \bf Acknowledgements} $\ $ It is a pleasure to thank Denis
Bernard for discussions.

\end{document}